\documentclass[twocolumn]{aastex62}

\shorttitle{Optical follow-up of three binary black hole mergers}
\shortauthors{Kim et al.}

\begin{document}

\title{GECKO Optical Follow-up Observation of Three Binary Black Hole Merger Events: GW190408\_181802, GW190412, and GW190503\_185404}

\correspondingauthor{Myungshin Im}
\email{mim@astro.snu.ac.kr}

\author{Joonho Kim}
\affil{Astronomy Program, Department of Physics and Astronomy, Seoul National University, Seoul 08826, Korea}
\affil{SNU Astronomy Research Center, Seoul National University, 1 Gwanak-ro, Gwanak-gu, Seoul 08826, Korea}

\author{Myungshin Im}
\affil{Astronomy Program, Department of Physics and Astronomy, Seoul National University, Seoul 08826, Korea}
\affil{SNU Astronomy Research Center, Seoul National University, 1 Gwanak-ro, Gwanak-gu, Seoul 08826, Korea}

\author{Gregory S. H. Paek}
\affil{Astronomy Program, Department of Physics and Astronomy, Seoul National University, Seoul 08826, Korea}
\affil{SNU Astronomy Research Center, Seoul National University, 1 Gwanak-ro, Gwanak-gu, Seoul 08826, Korea}

\author{Chung-Uk Lee}
\affil{Korea Astronomy and Space Science Institute, Daejeon 34055, Korea}

\author{Seung-Lee Kim}
\affil{Korea Astronomy and Space Science Institute, Daejeon 34055, Korea}

\author{Seo-Won Chang}
\affil{Astronomy Program, Department of Physics and Astronomy, Seoul National University, Seoul 08826, Korea}
\affil{SNU Astronomy Research Center, Seoul National University, 1 Gwanak-ro, Gwanak-gu, Seoul 08826, Korea}

\author{Changsu Choi}
\affil{Astronomy Program, Department of Physics and Astronomy, Seoul National University, Seoul 08826, Korea}
\affil{SNU Astronomy Research Center, Seoul National University, 1 Gwanak-ro, Gwanak-gu, Seoul 08826, Korea}

\author{Sungyong Hwang}
\affil{Astronomy Program, Department of Physics and Astronomy, Seoul National University, Seoul 08826, Korea}
\affil{SNU Astronomy Research Center, Seoul National University, 1 Gwanak-ro, Gwanak-gu, Seoul 08826, Korea}

\author{Wonseok Kang}
\affil{Deokheung Optical Astronomical Observatory, National Youth Space Center, Goheung 59567, Korea}

\author{Sophia Kim}
\affil{Astronomy Program, Department of Physics and Astronomy, Seoul National University, Seoul 08826, Korea}
\affil{SNU Astronomy Research Center, Seoul National University, 1 Gwanak-ro, Gwanak-gu, Seoul 08826, Korea}

\author{Taewoo Kim}
\affil{Deokheung Optical Astronomical Observatory, National Youth Space Center, Goheung 59567, Korea}

\author{Hyung Mok Lee}
\affil{Astronomy Program, Department of Physics and Astronomy, Seoul National University, Seoul 08826, Korea}
\affil{SNU Astronomy Research Center, Seoul National University, 1 Gwanak-ro, Gwanak-gu, Seoul 08826, Korea}
\affil{Korea Astronomy and Space Science Institute, Daejeon 34055, Korea}

\author{Gu Lim}
\affil{Astronomy Program, Department of Physics and Astronomy, Seoul National University, Seoul 08826, Korea}
\affil{SNU Astronomy Research Center, Seoul National University, 1 Gwanak-ro, Gwanak-gu, Seoul 08826, Korea}

\author{Jinguk Seo}
\affil{Astronomy Program, Department of Physics and Astronomy, Seoul National University, Seoul 08826, Korea}
\affil{SNU Astronomy Research Center, Seoul National University, 1 Gwanak-ro, Gwanak-gu, Seoul 08826, Korea}

\author{Hyun-Il Sung}
\affil{Korea Astronomy and Space Science Institute, Daejeon 34055, Korea}



\begin{abstract}
We present optical follow-up observation results of three binary black hole merger (BBH) events, GW190408\_181802, GW190412, and GW190503\_185404, which were detected by the Advanced LIGO and Virgo gravitational wave (GW) detectors. Electromagnetic (EM) counterparts are generally not expected for BBH merger events. However, some theoretical models suggest that EM counterparts of BBH can possibly arise in special environments, prompting motivation to search for EM counterparts for such events. We observed high-credibility regions of the sky for the three BBH merger events with telescopes of the Gravitational-wave EM Counterpart Korean Observatory (GECKO), including the KMTNet. Our observation started as soon as 100 minutes after the GW event alerts and covered 29 - 63 deg$^2$ for each event with a depth of $\sim$ 22.5 mag in $R$-band within hours of observation. No plausible EM counterparts were found for these events, but from no detection in the GW190503\_185404 event, for which we covered 69\% credibility region, we place the BBH merger EM counterpart signal to be $M_{g} > -18.0$ AB mag within about 1 day of the GW event. The comparison of our detection limits with light curves of several kilonova models suggests that a kilonova event could have been identified within hours from GW alert with GECKO observations if the compact merger happened at $< 400$ Mpc and the localization accuracy was of order of 50 deg$^2$. Our result gives a great promise for the GECKO facilities to find EM counterparts within a few hours from GW detection in future GW observation runs.
\end{abstract}



\section{Introduction} \label{sec:intro}
Since the first detection of the binary black hole (BBH) merger event by the Laser Interferometer Gravitational-Wave Observatory (LIGO) Collaboration and Virgo Collaboration (LVC) in 2015 September 14 \citep{2016PhRvL.116f1102A}, the reported number of BBH events has been increasing fast, from two in the the Advanced LIGO and Advanced Virgo (LIGO/Virgo) first (O1) observing run, and three in the second (O2) observing run, and to 20 in the first half (O3a) of the third (O3) observing run \citep{2020arXiv201014527A}. The gravitational wave (GW) astronomy is now flourishing with the GW signals providing valuable information on the compact star mergers such as the mass of the binary compact objects, and approximate distance and the location in the sky of the event.

The power of GW astronomy was much enhanced when the first electromagnetic-wave (EM) counterpart of a GW event was identified on 2017 August 17 during the LIGO/Virgo O2 run. The GW event, GW170817, was found to be due to a binary neutron star (BNS) merger with 90\% localization of 31 deg$^2$ and the luminosity distance of 40 Mpc \citep{2017PhRvL.119p1101A}. The combined efforts of the GW and EM observations produced a series of important results, including the first identification (AT2017gfo) of the exact location and the distance to the GW event \citep{2017ApJ...848L..12A, 2017Natur.551...64A, 2017Sci...358.1556C, 2017ApJ...850L...1L, 2017ApJ...848L..16S, 2017ApJ...848L..27T, 2017ApJ...848L..24V}, convincing observational evidence for the link between a short gamma-ray bursts (GRB) and a BNS merger \citep{2017ApJ...848L..13A, 2017ApJ...848L..14G, 2017ApJ...848L..15S}, undeniable discovery of a kilonova and the first detailed characterization of such an event \citep{2017PASA...34...69A, 2017Natur.551...64A, 2017ApJ...848L..19C, 2017ApJ...848L..17C, 2017Sci...358.1570D, 2017Sci...358.1565E, 2017ApJ...848L..25H, 2017Sci...358.1579H, 2017Sci...358.1559K, 2017ApJ...848L..20M, 2017ApJ...848L..18N, 2017Natur.551...67P, 2017Natur.551...75S, 2017Natur.551...71T, 2017PASJ...69..101U}, a glimpse on the BNS merger environment \citep{2017ApJ...849L..16I, 2017ApJ...848L..27T}, and the first application of the GW siren method to measure the Hubble constant \citep{2017Natur.551...85A}. In short, this was a monumental event that marked the beginning of multi-messenger astronomy (MMA) using GW and EM observations. If a similar study can be performed for BBH GW events that currently occupy the majority of the detected GW events, it would much broaden the power of MMA and enhance our knowledge on the previously rarely studied aspect of the universe.

As a result, many follow-up observations have been carried out to find EM counterparts of BBH GW events, yet no EM counterparts have been identified for the BBH events \citep{2016ApJ...826L..29, 2016PASJ...68L...9M, 2016ApJ...827L..40S, 2016MNRAS.462.4094S, 2016ApJ...823L..33S, 2017MNRAS.465.3656L, 2017ApJ...850..149S, 2017PASJ...69....9Y, 2019MNRAS.485.5180S, 2019ApJ...873L..24D, 2019ApJ...875...59Y, 2020A&C....3300425H, 2020MNRAS.497..726G, 2020MNRAS.492.1731G, 2020RAA....20...13T}. There are several reasons for the non detections. First of all, the strong gravitational force of black holes do not allow debris from the merger to escape, thus prohibiting EM-producing physical processes to occur \citep{2014ARA&A..52...43B}. A few models suggest that an EM counterpart might be produced if, e.g., the BBH merger preferentially happened in the active galactic nucleus (AGN) accretion disk, but the expected strength of EM radiation is weak or very uncertain \citep{2016ApJ...819L..21L, 2016ApJ...821L..18P, 2018MNRAS.477.4228P, 2017ApJ...835..165B, 2017ApJ...839L...7D, 2017MNRAS.464..946S, 2018ApJ...866...66M}. For the O3 run, a candidate of optical counterpart for BBH merger, GW190521 \citep{2020arXiv201014527A}, is reported as a possible AGN related transient \citep{2020PhRvL.124y1102G}. Secondly, the early LIGO/Virgo runs have been hampered by limited detector sensitivities, and as a result, the locations of the events estimated from the GW signals were very uncertain at a level of several hundreds to thousands deg$^{2}$ and their distance estimates were uncertain as well \citep{2019PhRvX...9c1040A}. Despite of these daunting prospects identification of BBH-induced GW EM counterparts remains attractive due to its potential for providing unexpected physical processes during the extreme compact merger and the prospects for using BBH merger EM counterparts for a number of MMA applications.

The LIGO/Virgo O3 run provided a unique opportunity to improve our limits of the BBH EM counterpart search. With the improved sensitivity of the GW detectors, the number of GW events increased (36 BBH events) and the localization area was improved as well (514 deg$^{2}$, median of 90\% localization area for BBH events\footnote{\url https://gracedb.ligo.org}). Hence we organized an EM follow-up observation group in Korea, utilizing a network of telescopes worldwide that were already in use for monitoring nearby galaxies for new transients such as supernova (Intensive Monitoring Survey of Nearby Galaxies, IMSNG; \citealt{2019JKAS...52...11I}). Our EM-follow-up effort was named as the Gravitational-wave EM Counterpart Korean Observatory (GECKO, \citealt{2020grbg.conf...25I}; Paek et al. 2021 in preparation). Several of the GECKO facilities have wide-field capabilities (field-of-view $>$ 1 deg$^{2}$), and suited for searching EM counterparts over a wide area.

In this paper, we present the GECKO optical follow-up observation for first three BBH merger events in the O3a run, namely GW190408\_181802 (hereafter, GW190408 for brevity), GW190412, and GW190503\_185404 (hereafter, GW190503). We focus our attention on the observation made by the Korea Microlensing Telescope Network (KMTNet; \citealt{2016JKAS...49...37K}), which is the most sensitive wide-field telescope of GECKO. The description of the KMTNet observation and analysis results will serve as a reference for future EM follow-up attempts. However, observations made by other telescopes are also reported to show the capabilities of GECKO facilities. In Section 2, we summarize three GW events which are targeted for our optical follow-up observation. Section 3 describes the telescopes we used and the observation strategy. Data analysis of the observed images is given in Section 4. We report the observation results and transient candidates in Section 5. Finally, we discuss the result in Section 6 and conclude the paper in Section 7.

\begin{deluxetable*}{lccccccccc}
\tablecaption{Target GW events\label{tab:table}}
\tablehead{
\colhead{Event} & \colhead{Type} & \colhead{Event time} & \colhead{Initial} & \colhead{Initial} & \colhead{GWTC-2} & \colhead{GWTC-2} & \colhead{FAR} \\[-1.5ex]
\colhead{} & \colhead{} & \colhead{[UT]} & \colhead{90\% Area [deg$^2$]} & \colhead{Distance [Mpc]} & \colhead{90\% Area [deg$^2$]} & \colhead{Distance [Mpc]} & \colhead{[yr$^{-1}$]}
}
\startdata
GW190408 & BBH & 2019-04-08 18:18:02 & 387 & 1473 $\pm$ 358 & 140 & $1580^{+400}_{-590}$ & $<$ 7.9 $\times 10^{-5}$ \\
GW190412 & BBH & 2019-04-12 05:30:44 & 156 & 812 $\pm$ 194 & 21 & $740^{+140}_{-170}$ & $<$ 7.9 $\times 10^{-5}$ \\
GW190503 & BBH & 2019-05-03 18:54:04 & 448 & 412 $\pm$ 105 & 94 & $1520^{+710}_{-660}$ & $<$ 7.9 $\times 10^{-5}$ \\
\enddata
\tablenotetext{}{Note. The FAR values are taken from the result of PyCBC BBH pipeline in the GWTC-2 catalog.}
\end{deluxetable*}

\section{Gravitational Wave Events} \label{sec:gw}
Here we describe the characteristics of three BBH events observed by GECKO in the early phase of the O3 run. We note that our observations were performed using the information from the initial alerts (bayestar.fits), and which were updated later in GWTC-2 catalog \citep{2020arXiv201014527A}.

\subsection{GW190408}
On 2019 April 8 18:18:02 UT, the first event of LIGO/Virgo O3 run, designated as originally as S190408a \citep{2019GCN.24069....1L}, and later as GW190408 \citep{2020arXiv201014527A}, was detected by the Advanced LIGO and Advanced Virgo GW detectors. This event was classified as a BBH merger with $>$ 99\% probability. The luminosity distance and the 90\% sky localization area were initially reported as 1473 $\pm$ 358 Mpc and 387 deg$^2$ (Table 1). Later, the further analysis of the GW data shows that the luminosity distance of $1580^{+400}_{-590}$ Mpc with a redshift of 0.30$^{+0.06}_{-0.10}$, the 90\% credible sky area of 140 deg$^{2}$, and a false alarm rate (FAR) of $<$ 7.9 $\times 10^{-5}$ yr$^{-1}$ \citep{2020arXiv201014527A}. The BH masses of the compact objects are found to be 24.5$^{+5.1}_{-3.4}\,M_{\odot}$ and 18.3$^{+3.2}_{-3.5} \, M_{\odot}$. The most recent GWTC-2 localization shows that the event happened in the northern hemisphere near RA = 351.0 deg and Dec = 53.9 deg, but the initial 90\% localization area contained a localized region in the southern hemisphere (Figure 1). Hence, the southern hemisphere localization area was observed by the KMTNet.

\subsection{GW190412}
GW190412, originally a candidate super-event S190412m, \citep{2019GCN.24098....1L}, was detected by three GW detectors on 2019 April 12 05:30:44 UT with $>$ 99\% probability of being BBH merger. Its 90\% credibility area on the sky and the luminosity distance were estimated as 156 deg$^{2}$ and 812 $\pm$ 194 Mpc (Table 1). These numbers were updated recently in the GWTC-2 catalog as 21 deg$^{2}$, $740^{+140}_{-170}$ Mpc, and FAR of $<$ 7.9 $\times 10^{-5}$ yr$^{-1}$ \citep{2020arXiv201014527A}. The redshift is given as 0.15$^{+0.03}_{-0.03}$ in the catalog. This was an event of special interest since this was the first BBH merger with definitely asymmetric component masses of 30.0$^{+4.7}_{-5.1} \,M_{\odot}$ and 8.3$^{+1.6}_{-0.9} M_{\odot}$, providing valuable information on the black hole spin, and revealing evidence for higher harmonics in the signal. The peak of GW event localization area is now estimated to be RA = 218.5 deg and Dec = 36.4 deg, but the initial estimates of the GW event location included an extensive region near equator, therefore this field was observed with GECKO facilities in the southern hemisphere as well (Figure 1).

\subsection{GW190503}
GW190503 was detected as a candidate event, S190503bf, on 2019 May 3 18:54:04 UT \citep{2019GCN.24377....1L}. With 96\% probability, this event was classified as a BBH merger. Initially, the 90\% localization region covered 448 deg$^2$ of the sky and the luminosity distance was estimated to be 412 $\pm$ 105 Mpc (Table 1). The most recent estimate in the GWTC-2 catalog improved the 90\% localization estimate to 94 deg$^{2}$, the distance estimate to $1520^{+710}_{-660}$ Mpc, a redshift as 0.29$^{+0.11}_{-0.11}$, and FAR of $<$ 7.9 $\times 10^{-5}$ yr$^{-1}$ \citep{2020arXiv201014527A}. We note that the distance estimate changed significantly from the initial values for this particular event. The BH masses of the binary system are found to be 42.9$^{+9.2}_{-7.8} \,M_{\odot}$ and 28.5 $^{+7.5}_{-7.9} M_{\odot}$ which are rather ordinary among the BBH mergers detected in the LVC runs. The 90\% localization area shrunk by a factor of 5 since the initial estimate, and the most recent estimate only partially overlaps with the initial area. The initial and the most up-to-date location estimates are all in the southern hemisphere, and it served as a prime target for the KMTNet follow-up observation.

\begin{figure*}
\centering
\includegraphics[width=180mm,angle=0]{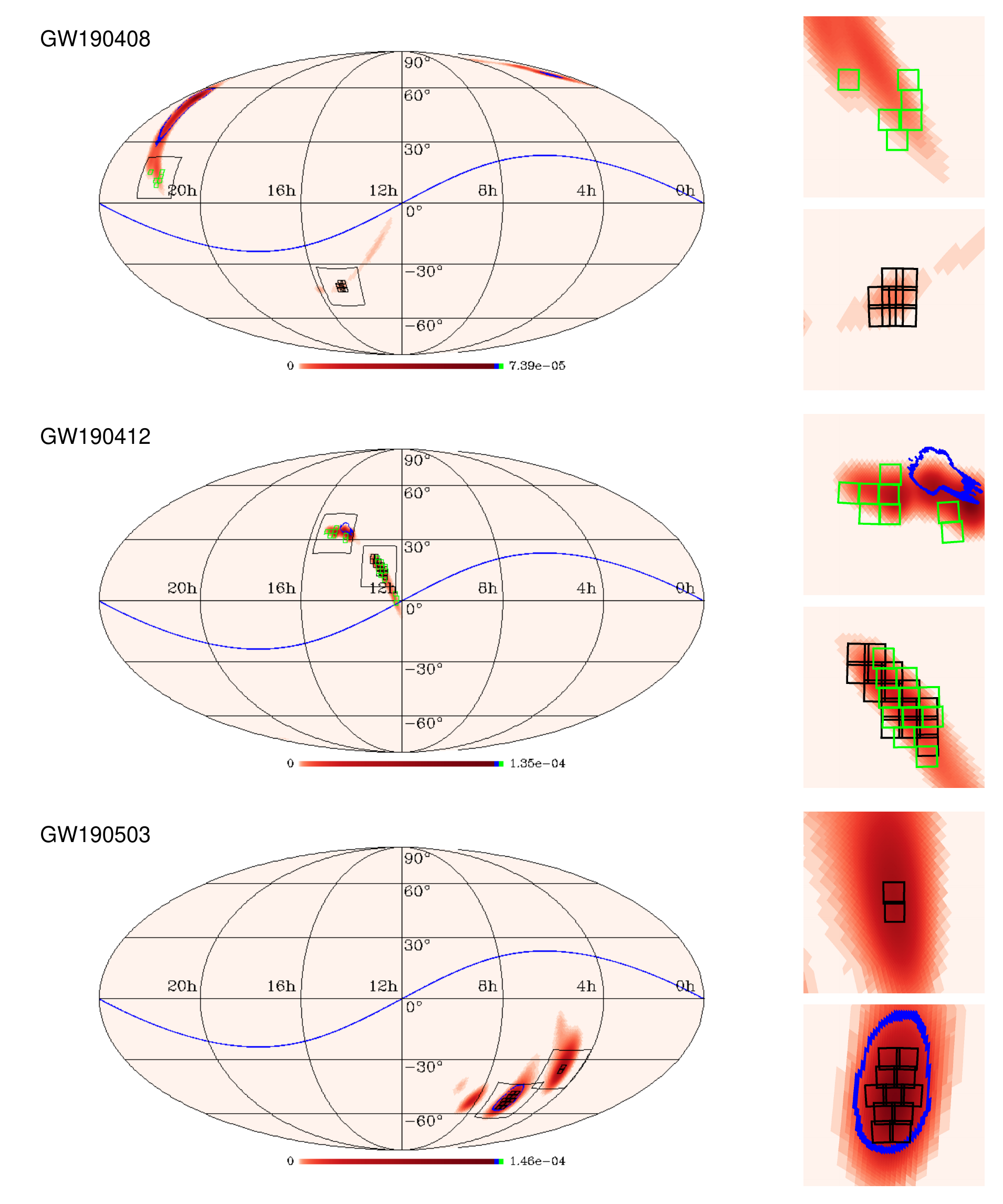}
\caption{The GW localization and the coverage of our EM follow-up tiling observation for three GW events. Here, we used the GW localization map of the initial alerts (bayestar.fits), while the blue contours show the boundary of the 90\% localization area in the GWTC-2 catalog. The large black boxes in the sky map on the left are expanded in the right panels. The small black and green boxes are the field coverages of the KMTNet and WIT pointings, respectively. The blue lines indicate the ecliptic plane.}
\end{figure*}

\section{Observation} \label{sec:obs}

\subsection{Telescopes}
For the follow-up observation of the GW events, we mainly used the KMTNet \citep{2016JKAS...49...37K} which is a network of three identical 1.6 m telescopes with a wide field of view ($\sim$ 4 deg$^2$) located at the Cerro Tololo Inter-American Observatory (CTIO), the South African Astronomical Observatory (SAAO), and the Siding Spring Observatory (SSO). The large field of view of the KMTNet enables us to cover the large GW localization area (See Section 2.3). Also, the three-site operation makes it possible to perform a follow-up observation 24-hrs a day when the target is observable (e.q. \citealt{2017Natur.551...71T, 2018JKAS...51...89K}). The KMTNet camera contains four 9k by 9k CCD chips, each covering about 1 deg$^{2}$. There is a gap of about $3\farcm1$ (east-west) to $6\farcm2$ (north-south) between each chip. To fill the CCD gaps, all the KMTNet observations were carried out by dithering between exposures.

Additionally, we used four telescopes of IMSNG \citep{2019JKAS...52...11I} to observe the northern hemisphere, especially at high declination regions ($\delta > 25^{\circ}$) that the KMTNet cannot reach. The wide-field IFU telescope (WIT), a wide-field ($\sim$ 5.5 deg$^2$) imager with a 0.25 m telescope equipped with multiple medium-band and broad-band filters \citep{2020arXiv201114049H} at the McDonald Observatory (McD) performed tiling observation to cover the wide GW localization area, but to a rather shallow depth ($\sim$19.0 mag in $V$-band at 5-$\sigma$ depth with a 7.5 min exposure). The other three are 1.0 m telescopes at the Deokheung Optical Astronomy Observatory (DOAO), the Mt. Lemmon Optical Astronomy Observatory (LOAO), and the Seoul National University Astronomical Observatory (SAO). These telescopes have a field of view of $13\farcm2$ x $13\farcm2$, $28\farcm1$ x $28\farcm1$, $21\farcm2$ x $21\farcm2$, respectively. These telescopes are suitable for observing galaxies that are likely to be the host galaxy of the BBH merger (targeted observation). The 5-$\sigma$ depth of the three 1.0 m telescopes are $\sim$ 19.5 mag in $R$-band with a 3 min exposure.

\subsection{Basic observation strategy}
Since there are both wide-field and narrow-field GECKO telescopes, we adopted two observing strategies. With wide-field telescopes (FoV $> 1$ deg$^{2}$), we performed tiling observation with an aim of covering the entire 90\% localization area. When the localization area is too large, the limit for the credibility of the localization is reduced from 90\% to a lower value. Also, when the localization area is several thousands deg$^{2}$, we switched to the target observation in the initial phase of GECKO observation as described below.

Mostly for the narrow-field telescopes but also for the case of the localization area is too large to cover even with a wide-field telescope, we made a prioritized list of galaxies that are likely to be host galaxies of the GW event. Galaxies are drawn from the GLADE catalog \citep{2018MNRAS.479.2374D} version 2.3, and we compute the likelihood score of a galaxy to be the host galaxy of the event. For both BNS and BBH events, the likelihood score implements the 3-D localization probability, and the $K$-band absolute magnitude. Here, $K$-band is chosen as a way to sort the list, since theoretical expectation is such that the more massive the host galaxy is, the more likely it hosts a BBH or BNS event \citep{2018MNRAS.481.5324M,2019MNRAS.487.1675A} and the $K$-band is known to be a good proxy for galaxy stellar mass (e.g., \citealt{2013ApJ...766..109K}).

To create a list of the observation coordinates for both the tiling and target observations, we established an automatic alert system which creates the list based on the GW event alert and the localization map. The coordinates are either handed over to the telescope operators or fed into the automatic observation planner. Exposure times were chosen to be typically 4 min per a tile for the KMTNet observation, and about 3 - 7.5 min for other telescopes. We use the observation of the three initial O3 events as test cases, and adjusted the exposure time.

More details on the galaxy prioritization and observational strategy will be presented in Paek et al. (2021, in preparation). 

\begin{deluxetable*}{lccccccc}
\tablecaption{Tiling observation of the three GW events\label{tab:table}}
\tablehead{\colhead{Event} & \colhead{Telescope} & \colhead{Start Time} & \colhead{Time Since} & \colhead{RA\tablenotemark{a}} & \colhead{Dec\tablenotemark{a}} & \colhead {Band} & \colhead{5-$\sigma$ Depth} \\[-1.5ex]
& & \colhead{[UT]} & \colhead{GW Alert [hours]} & \colhead{[hh:mm:ss]} & \colhead{[dd:mm:ss]} & \colhead{} & \colhead{[AB mag]}
}
\startdata
GW190408 & KMTNet & 2019-04-08 19:58:59 & 1.66 & 14:52:14.3 & -41:22:52 & R & 22.5 \\
GW190408 & WIT & 2019-04-09 11:32:26 & 17.2 & 21:43:19.9 & +15:04:05 & V & 17.6 \\
GW190412 & KMTNet & 2019-04-12 11:30:06 & 6.00 & 12:49:08.5 & +15:11:24 & R & 21.8 \\
GW190412 & WIT & 2019-04-12 07:17:53 & 1.78 & 12:13:33.8 & -00:57:38 & V & 18.3 \\
GW190503 & KMTNet & 2019-05-03 23:00:35 & 4.09 & 04:50:41.2 & -33:47:28 & R & 22.5 \\
\enddata
\tablenotetext{}{(This table is published in its entirety in the machine-readable format. A portion is shown here for guidance regarding its form and content.)}
\tablenotetext{a}{Central coordinate of the field.}
\end{deluxetable*}

\begin{deluxetable*}{lcccccc}
\tablecaption{Galaxy observation of GW190412\label{tab:table}}
\tablehead{\colhead{Telescope} &\colhead{Time} & \colhead{Name\tablenotemark{a}} & \colhead{RA} & \colhead{Dec} & \colhead {Band} & \colhead{5-$\sigma$ Depth} \\[-1.5ex]
& \colhead{[UT]} & & \colhead{[hh:mm:ss]} & \colhead{[dd:mm:ss]} & \colhead{} & \colhead{[AB mag]}
}
\startdata
DOAO & 2019-04-12 10:57:48 & PGC1631373 & 13:13:56.2 & +20:35:23 & $R$ & 20.5 \\
DOAO & 2019-04-12 11:08:47 & 2MSX J12252912+0317180 & 12:25:29.1 & +03:17:18 & $R$ & 20.3 \\
SAO & 2019-04-12 11:13:55 & PGC1631373 & 13:13:56.2 & +20:35:23 & $R$ & 18.1 \\
SAO & 2019-04-12 12:08:38 & 2MSX J12305009+0916524 & 12:30:50.1 & +09:16:52 & $R$ & 19.0 \\
LOAO & 2019-04-16 10:48:19 & PGC2053980 & 14:53:17.0 & +34:44:59 & $R$ & 19.4 \\
\enddata
\tablenotetext{}{(This table is published in its entirety in the machine-readable format. A portion is shown here for guidance regarding its form and content.)}
\tablenotetext{a}{Target name come from the GLADE catalog \citep{2018MNRAS.479.2374D}.}
\end{deluxetable*}

\begin{deluxetable*}{lccccccccc}
\tablecaption{Observation summary\label{tab:table}}
\tablehead{\colhead{Event} & \colhead{Telescope} & \colhead{Observation start} & \colhead{Observation end} & \colhead{N\tablenotemark{a}} & \colhead{Covered Area} & \colhead{Band} & \colhead{5-$\sigma$ Depth [AB mag]} 
}
\startdata
GW190408 & KMTNet & 2019-04-08 19:58:59 & 2019-04-09 21:54:00 & 8\tablenotemark{b} & 29 deg$^2$ & $BR$ & 21.5 - 23.3 \\
GW190408 & WIT & 2019-04-09 11:32:26 & 2019-04-12 11:34:12 & 6\tablenotemark{b} & 33 deg$^2$ & $V$ & 17.1 - 18.7 \\
\hline
GW190412 & SAO & 2019-04-12 11:13:55 & 2019-04-14 19:29:56 & 76\tablenotemark{c} & & $R$ & 18.0 - 20.7 \\
GW190412 & WIT & 2019-04-12 07:17:53 & 2019-04-14 11:35:23 & 22\tablenotemark{b} & 116 deg$^2$ & $V$ & 17.6 - 19.2 \\
GW190412 & DOAO & 2019-04-12 10:57:48 & 2019-04-17 15:11:07 & 81\tablenotemark{c} & & $BVRI$ & 15.8 - 21.5 \\
GW190412 & KMTNet & 2019-04-12 11:30:06 & 2019-04-13 13:10:13 & 15\tablenotemark{b} & 63 deg$^2$ & $R$ & 20.2 - 22.9 \\
GW190412 & LOAO & 2019-04-16 10:48:19 & 2019-04-16 11:09:56 & 2\tablenotemark{c} & & $V$ & 19.3 - 19.4 \\
\hline
GW190503 & KMTNet & 2019-05-03 23:00:35 & 2019-05-05 00:17:17 & 13\tablenotemark{b} & 52 deg$^2$ & $R$ & 21.8 - 22.8 \\
\enddata
\tablenotetext{}{${}^{a}$Number of the observed targets. ${}^{b}$Tiling observation. ${}^{c}$Galaxy targeted observation.}
\end{deluxetable*}

\subsection{Observations}
\subsubsection{GW190408}
The first event, GW190408, was localized mostly in the northern hemisphere of the sky, but the area was too close to the Sun to observe. But, there was a 7 deg$^2$ of the initial 90\% localization area in the southern hemisphere around declination of $\sim$ -40 deg. We targeted this field using the KMTNet with 4 minutes exposure time in $B$ and $R$-bands. We started the observation about 100 minutes after the GW event alert, and we obtained four epochs ($\sim$ 8 hours interval) of observation data from three KMTNet sites. A portion of the northern hemisphere localization area of initial GW alert was observed with the WIT for three nights (2019 Apr 09 - 11). The observation covered a 32.6 deg$^2$ of the sky (6 fields) with 7.5 minutes exposure time in $V$-band. No target observation was performed since the event occurred at a distance that was too far to be covered by the GLADE catalog. Total probability region covered by the two telescopes is 2.0\% for the initial GW localization, and the probability region decreased to 0.14\% with respect to the GWTC-2 localization map. The reduction in the covered probability region from the initial to GWTC-2 localization suggests that EM follow-up observations may miss optical counterparts due to inaccuracy in the GW event localization. Improving the accuracy of rapid localization of GW events is important for the success of future optical follow-up efforts. The KMTNet and WIT observations are presented in Table 2, along with their 5-$\sigma$ point source detection limits.

\subsubsection{GW190412}
Most of the initial 90\% localization area (156 deg$^2$) of GW190412 lie on the northern hemisphere of the sky, at declination of $-$3 deg to +38 deg. It can be divided into two regions one at declination $<$ +25 deg (area1) and another at declination $>$ +25 deg (area2). The area1 was observable with the KMTNet and we started the observation at 6 hours after the event alert. We obtained the KMTNet images for a 63 deg$^2$ on the sky by tiling observation of 15 fields with 4 minutes exposure time in $R$-band at 2 epochs of 1 day interval. The high priority regions of both area1 and area2 were observed using the WIT by tiling 22 fields (116.3 deg$^2$) with a 7.5 min exposure time in $V$-band on April 12th and 14th, 2019. We also performed the target observation of host galaxy candidates with the narrow field telescopes. The target galaxies were chosen as described in Section 3.2. We observed 81, 76, and 2 galaxies with DOAO 1 m, SAO 1 m, LOAO 1 m telescopes, respectively, with exposure times of 3 - 5 minutes in $R$-band. Overall, we covered a 40\% probability region of the initial GW localization map, and added imaging data for 86 galaxy candidates. However, the covered probability region reduces to 2.3\% with respect to the GWTC-2 localization map because the 90\% region moved to declination $>$ +30 deg where our coverage is deficient (Figure 1). Like GW190408, the reduction is significant in the covered localization probability area from the initial to the GWTC-2 credible regions, reiterating the importance of rapid, accurate GW localization. The list of the tiling observation coordinates and the list of the observed host galaxy candidates are given in Tables 3. Also given are 5-$\sigma$ point source detection limits of the data.

\subsubsection{GW190503}
We performed follow-up observation for GW190503 only with the KMTNet because the initial 90\% credibility area was located in the southern hemisphere, the area was extensive (448 deg$^2$) and located in the southern hemisphere. Furthermore, the KMTNet could reach deepest among our facilities available in the southern hemisphere where another available GECKO facility was a 0.43 m Lee Sang Gak Telescope \citep{2015JKAS...48..207I, 2017JKAS...50...71C}. We observed a high probability area from the initial alert with 13 tiles with 4 minutes exposure time in $R$-band at two epochs separated by 1 day. The covered area is 52 deg$^2$, which corresponds to a 33\% probability region. Since the localization area has shrunk to 96 deg$^2$ in the GWTC-2 catalog, it turned out that the KMTNet observation covered the 69\% localization probability region of GWTC-2. Table 2 shows the central coordinates of the KMTNet pointings and their 5-$\sigma$ point source detection limits.

We summarize all of the GECKO observations in Table 4 and the area covered by the tiling observation is marked as rectangles in Figure 1.

\begin{figure*}
\centering
\includegraphics[width=120mm,angle=0]{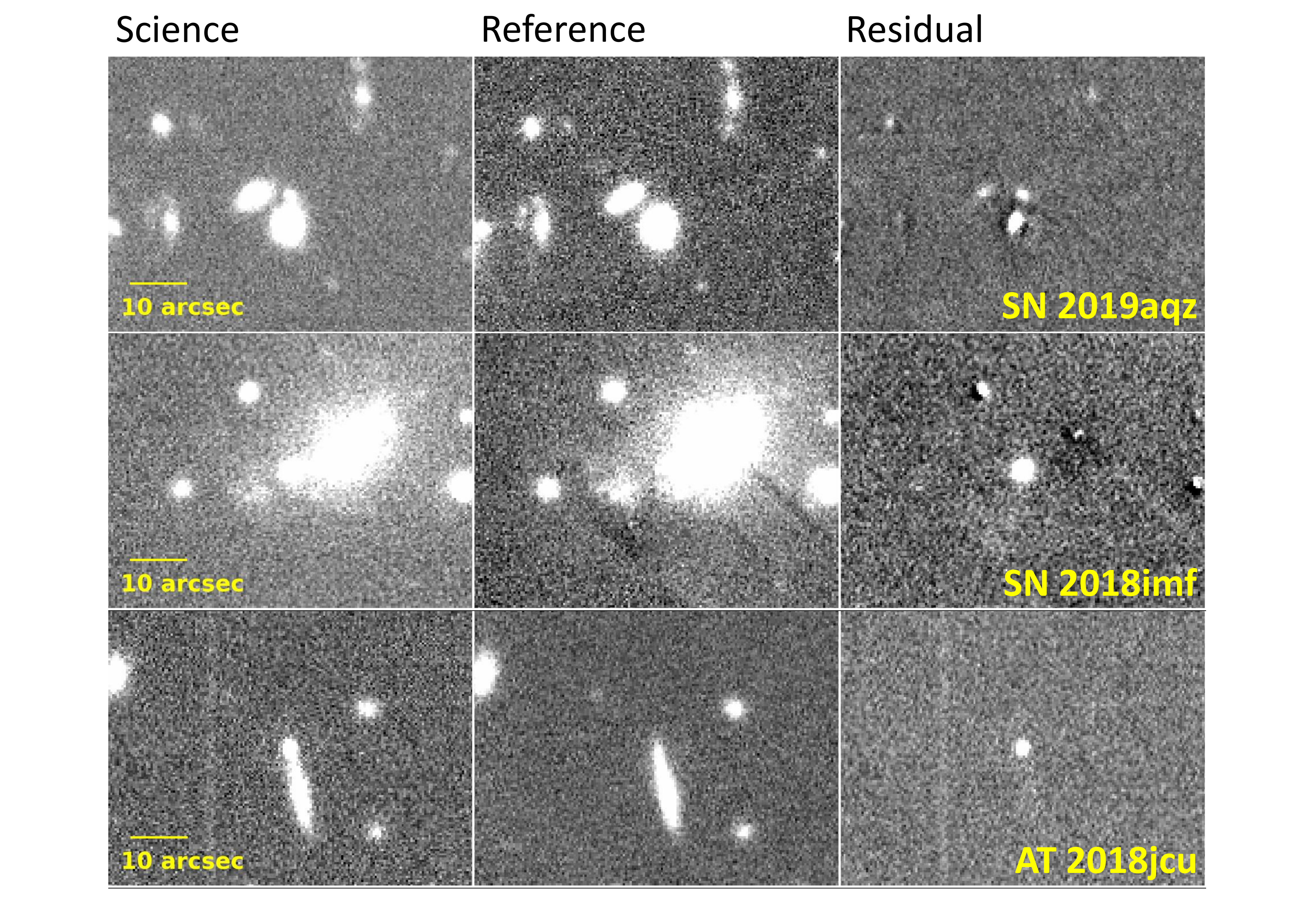}
\caption{Three transients found in the follow-up observation of GW190412 event. The top, middle, and bottom panels show SN 2019apz, SN 2018imf, and AT 2018jcu, respectively. From left to right, each panel represents the object (science) image, the reference image, and the residual image after subtracting the reference image from the object image.}
\end{figure*}

\section{Data Analysis and Transient Search} \label{sec:analysis}
All the observed images were processed with basic reduction procedures (bias subtraction, dark subtraction, and flat fielding). Astrometric calibration is done using SCAMP \citep{2006ASPC..351..112B} or Astrometry.net \citep{2010AJ....139.1782L}, using ten to hundreds stars as astrometry references depending on the field of view and depth. This produces images with astrometric accuracy of rms scatter of $0\farcs2 - 0\farcs8$ in both RA and Dec directions. For the KMTNet observation for a given epoch, only two frames were taken, which made it challenging to remove cosmic-rays during co-addition. Therefore, we removed the cosmic rays from each single frame using LA-Cosmic software \citep{2001PASP..113.1420V}. The sequentially observed two images were combined to produce a deeper image with SWarp \citep{2002ASPC..281..228B}.

We used Pan-STARRS (PS1, \citealt{2016arXiv161205560C}) images at the closest wavelength to the observed image filter as reference images (for example, use PS1 $r$-band image for KMTNet $R$-band image), and the PS1 reference images were subtracted from observed images are by convolving and re-scaling them to match the spatial resolution and the flux scale of the observed images using High Order Transform of PSF ANd Template Subtraction (HOTPANTS, \citealt{2015ascl.soft04004B}).

The declinations of the KMTNet images for GW190408 and GW190503 are lower than $-$30 deg, and out of the PS1 coverage. For these images, we observed the same area of the sky twice with a time separation of about 1 day, and we used the images observed $\sim$ 24 hours after the first epoch as reference images for the two events.

Transient candidates were detected by identifying positive signals having signal-to-noise ratio greater than five in the subtracted images by running SExtractor \citep{1996A&AS..117..393B}. Many of the detected sources in the residual images were false identification found at over or under-subtracted bright sources. These imperfect subtraction was often caused by a slight mismatch in astrometry and/or in the wavelength coverage of the reference and the observed images, and they appear as positive signals right next to negative pixels (see the upper part of the residual image of SN 2018imf in Figure 2). To remove these artifacts, we ran SExtractor on the inverse images of the residual images. The positive detections in the residual images were rejected if there are positive detections in the inverse residual image right next to their position ($< 3\arcsec$). Additionally, we removed candidates matched with non-transient sources in the reference images to exclude variable sources like variable stars or active galactic nuclei. We also removed candidates that are on bad pixels or whose central pixels were saturated. The remaining candidates were visually inspected and classified as new transient candidates if there were no other artifacts. Figure 2 shows the science, reference, and residual images of three transients found in the KMTNet images (see Section 5).

For the images from the other telescopes, we subtracted images using PS1 images and visually inspected the residual images to find transient candidates.

\begin{deluxetable*}{lccccccccc}
\tablecaption{List of detected transients and transient candidates\label{tab:table}}
\tablehead{
\colhead{Number} & \colhead{Observation time} & \colhead{RA} & \colhead{Dec} & \colhead{$R$ mag [AB]} & \colhead{Name}
}
\startdata
1 & 2019-04-12 11:42:31 & 12:49:11.59 & 14:08:00.25 & 21.34 $\pm$ 0.08 \\
2 & 2019-04-12 12:00:53 & 12:52:38.01 & 19:04:04.33 & 20.80 $\pm$ 0.07 \\
3 & 2019-04-12 12:57:24 & 12:36:45.09 & 11:43:00.09 & 21.06 $\pm$ 0.09 \\
4 & 2019-04-12 12:57:24 & 12:42:19.38 & 01:17:22.02 & 22.05 $\pm$ 0.13 \\
5 & 2019-04-12 12:57:24 & 12:45:07.81 & 10:37:10.78 & 21.33 $\pm$ 0.09 \\
6 & 2019-04-12 12:57:24 & 12:38:11.48 & 10:19:58.89 & 20.88 $\pm$ 0.05 & SN 2019aqz \\
7 & 2019-04-12 13:03:38 & 12:37:34.85 & 13:16:49.31 & 21.17 $\pm$ 0.06 \\
8 & 2019-04-12 13:03:38 & 12:37:42.61 & 13:20:19.88 & 21.40 $\pm$ 0.09 \\ 
9 & 2019-04-12 13:03:38 & 12:36:24.06 & 13:22:36.69 & 21.87 $\pm$ 0.10 \\
10 & 2019-04-12 13:03:38 & 12:41:25.46 & 13:21:30.02 & 21.24 $\pm$ 0.06 \\
11 & 2019-04-12 13:03:38 & 12:42:41.39 & 13:15:54.89 & 17.82 $\pm$ 0.004 & SN 2018imf \\
12 & 2019-04-12 13:03:38 & 12:43:54.16 & 13:12:12.77 & 21.15 $\pm$ 0.06 \\
13 & 2019-04-12 13:03:38 & 12:40:05.16 & 13:07:58.24 & 19.56 $\pm$ 0.02 & AT 2018jcu \\
\enddata
\end{deluxetable*}

\begin{deluxetable*}{lccccccccc}
\tablecaption{List of moving objects\label{tab:table}}
\tablehead{
\colhead{Number} & \colhead{Observation time} & \colhead{RA} & \colhead{Dec} & \colhead{$R$ mag [AB]} & \colhead{KMTNet Site}
}
\startdata
1 & 2019-04-12 12:57:24 & 12:42:02.43 & 11:32:13.17 & 20.38 $\pm$ 0.07 & SAAO \\
& 2019-04-12 13:07:08 & 12:42:02.03 & 11:32:12.56 & 20.55 $\pm$ 0.07 & SAAO \\
2 & 2019-04-12 12:51:10 & 12:36:30.73 & 14:09:43.95 & 21.68 $\pm$ 0.20 & SAAO \\
& 2019-04-12 13:03:38 & 12:36:31.07 & 14:09:40.43 & 21.72 $\pm$ 0.17 & SAAO \\
3 & 2019-04-12 13:03:38 & 12:40:39.09 & 13:46:32.90 & 21.62 $\pm$ 0.16 & SAAO \\
& 2019-04-12 13:13:15 & 12:40:38.80 & 13:46:33.75 & 21.56 $\pm$ 0.15 & SAAO\\ 
\enddata
\end{deluxetable*}

\section{Transient Search Result} \label{sec:result}
After the identification of the credible transient candidates, we matched them with known solar system objects using SkyBoT \citep{2006ASPC..351..367B} version 3.0. SkyBot is a system that provides the location of known solar system objects at a given epoch. Also we matched our transient candidates with those that were already reported in the Transient Name Server (TNS\footnote{\url https://wis-tns.weizmann.ac.il}).

For GW190408, we found two transient candidates. They are matched with known solar system objects.

In the observed areas for the GW190412 event, we detected 275 transient candidates. As we describe below, the majority of them are found to be the solar system objects, since the localization area of GW190412 was near the Ecliptic plane with an ecliptic latitude of about 14 deg. Among these candidates, 258 objects were found to have matches with solar system objects within 1 arcmin of their positions (most of the sources are matched within $5\arcsec$). In addition, we found three moving objects that did not have matching solar system objects. We consider these to be previously unknown solar system objects or known solar system objects with high positional errors. One transient candidate was found to be a fast proper motion star of $\sim$ $1\arcsec$/year \citep{2018A&A...616A...1G}. Three objects were found in the vicinity of possible host galaxies and they are identified in the TNS as SN 2019aqz (Type Ia), SN 2018imf (Type IIP) and AT 2018jcu (Figure 2). The remaining 10 candidates do not have a possible host galaxy in their vicinity, and only observed in a single frame. Their nature cannot be judged from our data alone, but we suspect that most of them are solar system objects considering the large number of such objects in this field. The observing strategy at that time was geared toward minimizing the overhead by reducing the time required to move to another field. However, we realize that  efficient identification of moving objects can be done more easily if subsequent frames were taken about an hour later. We are now considering such a strategy for future observations.

No transient candidates were found in the observed areas of GW190503 event. Also, no transient candidates were found in WIT, DOAO, SAO, and LOAO images in all the events.

Table 5 shows the three transients and 10 transient candidates with uncertain nature from the GECKO GW190412 observation. In Table 6, we also report the epochs and the positions of the solar system objects found from our follow-up observations.

\section{Discussion} \label{sec:discuss}

\subsection{Limits on BBH EM Counterpart Brightness}
As described above, we could not identify any plausible EM counterparts to the three BBH merger events. This is not very surprising, considering that BBH merger events are not expected to produce EM counterparts. Also, in light of the recently updated localization area (See Section 2), the KMTNet observation missed the updated area for GW190408 and GW190412, so identification of EM counterparts was not possible from our observation in retrospect. For GW190503, the KMTNet covered most of the updated 90\% credible region. No transients were found in this data, which places a useful upper limit on the brightness of the EM counterpart. We note, however, that the updated distance to GW190503 is $1527 \pm 411$ Mpc, so the EM counterpart must have been bright in order to be detected with the KMTNet observation. Below, we discuss the feasibility of detecting EM counterparts and upper limits that we could have imposed on these events if our observations covered all of the localization area.

\begin{figure*}
\centering
\includegraphics[width=200mm,angle=270]{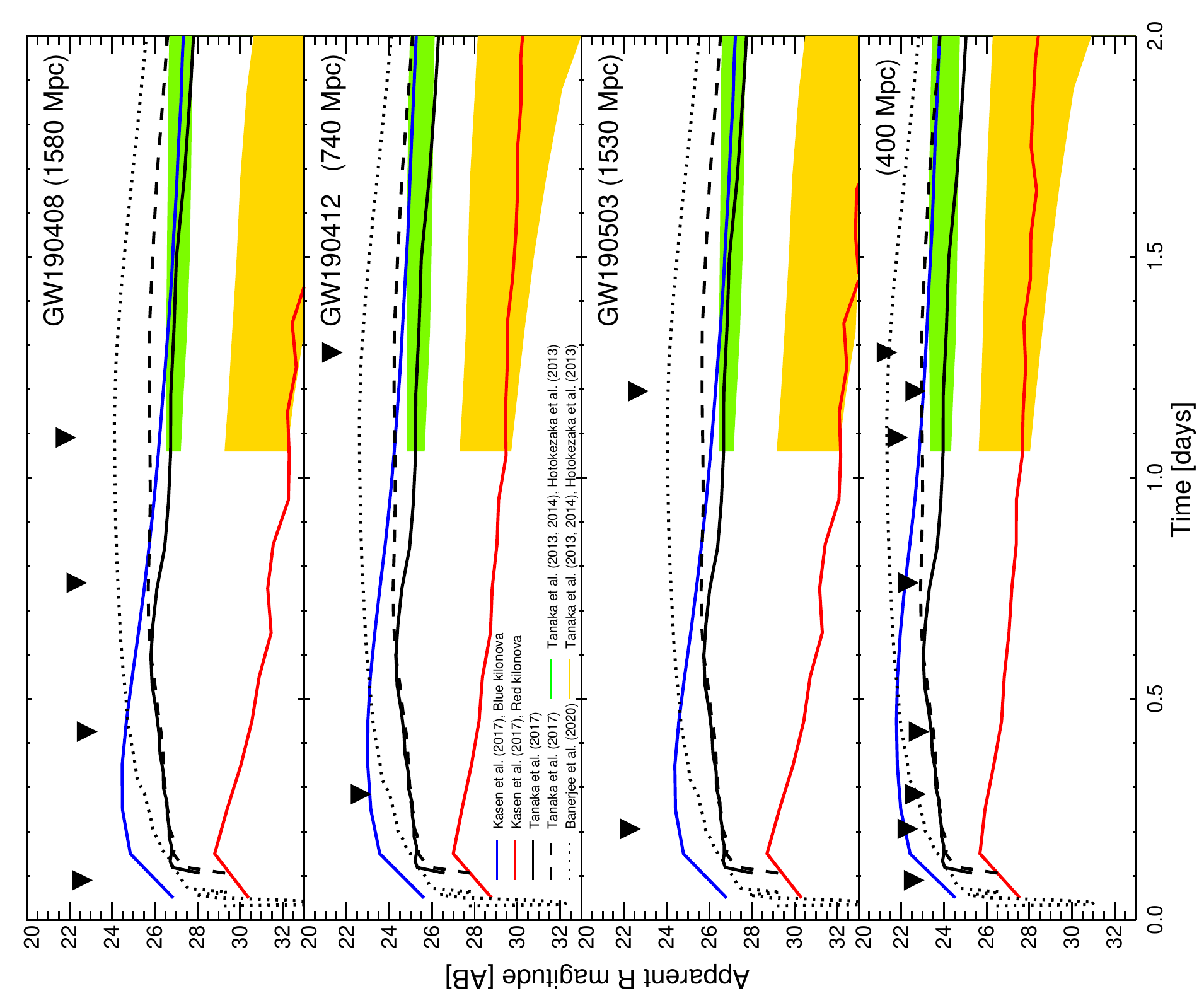}
\caption{Comparison of different model light curves of kilonova and the KMTNet image depths (black triangles) of three events in top three panels. The bottom panel assumes a kilonova distance of 400Mpc, and the KMTNet image depths from the observation of the three events are shown together. The blue and red solid lines indicate the blue and red components of the AT2017gfo-like kilonova model light curves of \citet{2017Natur.551...80K}. The black solid, dashed, and dotted lines are the AT2017gfo-like kilonova model of \citet{2017PASJ...69..102T}, and \citet{2020ApJ...901...29B}. These models assume ejecta masses of 0.03 $M_{\odot}$, 0.03 $M_{\odot}$, and 0.05 $M_{\odot}$ and electron fractions of 0.1 - 0.4, 0.25, and 0.3 - 0.4, respectively. The green and yellow regions are the range of four NSBH and five BNS model light curves, respectively, from \citet{2013ApJ...775..113T, 2014ApJ...780...31T, 2013ApJ...778L..16H}.}
\end{figure*}

From non-detection of transients for GW190503, we place an upper limit on the brightness of EM optical counterpart for the BBH event. The median 5-$\sigma$ depth of the KMTNet observations is about $R=22.5$ AB mag. Adopting the distance of 1500 Mpc, we get an upper limit for the EM counterpart brightness of $M_{R} \sim -18.3$ AB mag without K-correction. Since GW190503 occurred at $z \sim 0.32$ and the $R$-band has an effective wavelength of $\sim 650$ nm (e.g., \citealt{2010JKAS...43...75I}), the rest-frame wavelength covered by $R$-band for GW190503 is $\sim 490$nm, which is approximately the effective wavelength of $g$-band. Therefore, $g$-band K-correction is dominated by the bandwidth dilation term of $-2.5 \times \rm{log} (1+z)$ (e.g., \citealt{2020ApJ...897..163T}). Adopting $-2.5 \times \rm{log} (1+z)$ as the K-correction, we get a lower limit on $M_{g}$ (the absolute magnitude in $g$-band) of $M_{g} > -18.0$ AB mag (5-$\sigma$). This is about 1.5 to 2.5 magnitude brighter than the $g$-band peak magnitudes of various kilonova model light curves \citep{2013ApJ...775..113T, 2014ApJ...780...31T, 2017PASJ...69..102T, 2013ApJ...778L..16H, 2017Natur.551...80K, 2018ApJ...855L..23A, 2020ApJ...901...29B}.

\subsection{Comparison with kilonova light curve models}
We compare model light curves of kilonova and the imaging depths to discuss the possibility of GW EM counterpart detection (Figure 3). One of the models is the AT2017gfo-like kilonova model of \citet{2017Natur.551...80K}, hereafter K17. The model is for the AT2017gfo that combines the blue and red kilonova components with ejecta masses of 0.025 $M_{\odot}$ and 0.04 $M_{\odot}$, respectively. We plotted the two kilonova components in Figure 3 and it shows that the blue kilonova is more dominant than red kilonova in the early phase, $<$ 2 days. Similarly, the AT2017gfo-like kilonova models of \citet{2017PASJ...69..102T} and \citet{2020ApJ...901...29B} are plotted in Figure 3 (with ejecta masses of 0.03 - 0.05 $M_{\odot}$). We also plot earlier kilonova light curve predictions for NS-BH (NSBH) binary merger and BNS mergers with ejecta masses of 0.0007 - 0.07 $M_{\odot}$ \citep{2013ApJ...775..113T, 2014ApJ...780...31T, 2013ApJ...778L..16H}. These models assume a higher mass fraction of lanthanides ($X_{lan} \sim 10^{-1.5}$) than the K17 model ($X_{lan} = 10^{-4}$), and consequently predict fainter light curves than the AT2017gfo-like models. For the distances of these models for each event, we assumed the values from the GWTC-2 catalog. Overall, we confirm that these events are too far (700 - 1600 Mpc) for detecting their EM counterparts with KMTNet at a nominal depth. But if AT2017gfo-like kilonovae appeared at about 400 Mpc or less, KMTNet could have detected them (Figure 3).

\subsection{Prospects for rapid follow-up observation}
The early light curve can discern different mechanisms proposed for the observed blue optical-NIR emission in the early phase of EM counterpart of GW170817. The cooling of ejecta shock-heated by various mechanisms (e.g., cocoon model) would produce the early light curve that peaks very quickly in about $\sim$ hour, while the early light curve that is dominated by the radioactive decay would peak at in about a day \citep{2017Natur.551...64A}. Therefore, a rapid EM follow-up observation is important for constraining different EM production mechanisms.

For the GW190408 event, we were able to start the follow-up observation after 100 minutes from the GW alert. This demonstrates that fast follow-up observation is possible with our GECKO network, within about 1 hour from the GW event alert. Yet, it is necessary to shorten the start time of the follow-up observation, since it requires a long time even for wide-field telescopes such as KMTNet to cover the whole localization area of the GW events. The areas covered by our observations are 29, 63, and 52 deg$^2$ for GW190408, GW190412, and GW190503, respectively. From these data, we estimate that it took 1.69 minutes per deg$^2$ on average. For BNS events in the 4th (O4) GW observing run, the median 90\% credibility region is expected to be $33^{+5}_{-5}$ deg$^2$ \citep{2020LRR....23....3A}. Therefore, it would take about 50 minutes to cover the median 90\% localization areas of the O4 BNS events with a KMTNet tiling observation.

For the GECKO facilities to catch the early light curve of EM counterparts within an hour time-scale during the future O4 run, it is necessary to start the observation almost immediately after the initial GW alert. The work is in progress to shorten the latency of the starting time to 10 minutes or less. 

For all the events, we also note that the localization area shrunk significantly or the peak of the highest probability localization area moved to another location. Considering a large amount of time needed to cover a wide localization area, improvements in the accuracy of the initial localization map is highly desirable. If not, one should cover as much 90\% localization area as possible in order not to miss EM counterparts.

\subsection{Need for deep reference images}
When searching for transients through the image subtraction technique, we found that the PS1 $r$-band images could serve as acceptable reference images even if the filter system was slightly different from our $R$-band. However, during the analysis of the GW190408 and GW190503 optical data, we also discovered the lack of suitable reference images at Dec $< -30$ deg where there is no PS1 coverage, and the SkyMapper images \citep{2019PASA...36...33O} are rather shallow. The Dark Energy Survey (DES; \citealt{2018ApJS..239...18A}) images are available, covering 5000 deg$^2$ of the southern hemisphere with a depth of $\sim$ 24 mag. However the DES images provide only a partial coverage at $-$60 $<$ Dec (deg) $<$ $-$30. The sky below Dec = $-$60 deg will remain uncovered by deep imaging observations, until the Legacy Survey of Space and Time (LSST, \citealt{2019ApJ...873..111I}) starts its full operation. Therefore, we are now performing a wide imaging survey named the KMTNet Synoptic Survey of Southern Sky (KS4) to provide reference images for the KMTNet follow-up data to the depths of $\sim 23.5$ AB mag over 7000 deg$^{2}$ area at the Dec $< -60$ deg areas and some of the $-60 < \rm{Dec (deg)} < -30$ areas that are not deeply covered by previous surveys. The KS4 will allow us to quickly identify EM counterparts so that the EM counterparts can be studied in depth.

\section{Summary} \label{sec:sum}
In this paper, we presented the GECKO follow-up observation of three BBH events, GW190408, GW190412, and GW190503, that were reported early in the O3a run. In particular, we reported the KMTNet observations in detail which started at 100 minutes to 6 hours from the GW detection with a sky coverage of 29 - 63 deg$^2$ and $\sim$ 22.5 mag depth in $R$-band.

From our observation, we identified 13 transient candidates for GW190412 and hundreds of moving objects such as asteroids. Among these transient candidates, three transients were already reported (two confirmed as a supernova) and the remaining 10 are likely to be moving objects due to their proximity to the field near the ecliptic plane, the lack of a host galaxy near them, and being observed only once. Therefore, no probable EM counterpart was found in our observation. The improved localization of these events in the recently published GWTC-2 catalog reveals that the initial position estimates were very broad and many of the optical follow-up can point to the sky that are not in the improved localization area. However, we find that our KMTNet imaging observation covered a 69\% probability region of GWTC-2 localization map for GW190503, and our non-detection places an upper limit on the rest-frame $g$-band magnitude of EM counterpart of this BBH merger event to be $M_{g} > -18.0$ AB mag. Together with this, the comparison between the image depths of KMTNet data and the theoretical kilonova light-curves show that brighter AT2017gfo-like kilonova events can be detected when the luminosity distance is out to $\sim$ 400 Mpc.

Our follow-up observation started as quickly as about 100 minutes after the GW event alert, and could cover an area of about 50 deg$^2$ in one hour, demonstrating that the observation of the 90\% localization area in the future O4 run can be done in about two hours. The work is ongoing to improve the response time. Overall, our results show promises that GECKO can identify kilonova events out to 400 Mpc within hours from the GW event alert in future GW observing runs that will improve the detector sensitivity event localization capability.

\acknowledgments
We thank the referee for useful comments that helped and improved the presentation of the paper. We appreciate the support and encouragement from colleagues of the Korean Gravitational Wave Group. This research has made use of the KMTNet system operated by the Korea Astronomy and Space Science Institute (KASI) and the data were obtained at three host sites of CTIO in Chile, SAAO in South Africa, and SSO in Australia. We also thank the operators at LOAO and KMTNet for performing the requested observations and the staffs at the McDonald observatory for their support of WIT. This work was supported by the National Research Foundation of Korea (NRF) grant, No. 2020R1A2C3011091, funded by the Korea government (MSIT), and the Korea Astronomy and Space Science Institute under the R \& D program (Project No. 2020-1-600-05) supervised by the Ministry of Science and ICT. 

%

\vspace{5mm}
\facilities{KMTNet \citep{2016JKAS...49...37K}, WIT, DOAO, SAO, LOAO \citep{2019JKAS...52...11I}}


\software{SExtractor \citep{1996A&AS..117..393B}, 
          Scamp \citep{2006ASPC..351..112B},
          SWarp \citep{2002ASPC..281..228B}, 
          Astronomy.net \citep{2010AJ....139.1782L}, 
          SkyBoT (v3.0; \citealt{2006ASPC..351..367B}),
          HOTPANTS \citep{2015ascl.soft04004B},
          LA-Cosmic \citep{2001PASP..113.1420V}
          }


\begin{thebibliography}{}
\bibitem[Abbott et al.(2016)]{2016PhRvL.116f1102A} Abbott, B. P., Abbott, R., Abbott, T. D., et al.\ 2016, \prl, 116, 061102
\bibitem[Abbott et al.(2017a)]{2017PhRvL.119p1101A} Abbott, B. P., Abbott, R., Abbott, T. D., et al.\ 2017, \prl, 119, 1101
\bibitem[Abbott et al.(2017b)]{2017ApJ...848L..12A} Abbott, B. P., Abbott, R., Abbott, T. D., et al.\ 2017, \apjl, 848, 12
\bibitem[Abbott et al.(2017c)]{2017ApJ...848L..13A} Abbott, B. P., Abbott, R., Abbott, T. D., et al.\ 2017, \apjl, 848, 13
\bibitem[Abbott et al.(2017d)]{2017Natur.551...85A} Abbott, B. P., Abbott, R., Abbott, T. D., et al.\ 2017, \nat, 551, 85
\bibitem[Abbott et al.(2018)]{2018ApJS..239...18A} Abbott, T. M. C., Abdalla, F. B., Allam, S., et al.\ 2018, \apjs, 239, 18
\bibitem[Abbott et al.(2019)]{2019PhRvX...9c1040A} Abbott, B. P., Abbott, R., Abbott, T. D., et al.\ 2019, \prl, 9, 1040
\bibitem[Abbott et al.(2020a)]{2020arXiv201014527A} Abbott, R., Abbott, T. D., Abraham, S. et al.\ 2020, arXiv:2010.14527
\bibitem[Abbott et al.(2020b)]{2020LRR....23....3A} Abbott, B. P., Abbott, R., Abbott, T. D., et al.\ 2020, LRR, 23, 3
\bibitem[Andreoni et al.(2017)]{2017PASA...34...69A} Andreoni, I., Ackley, K., Cooke, J., et al.\ 2017, \pasa, 34, 69
\bibitem[Arcavi et al.(2017)]{2017Natur.551...64A} Arcavi, I., Hosseinzadeh, G., Howell, D. A., et al.\ 2017, \nat, 551, 64
\bibitem[Arcavi(2018)]{2018ApJ...855L..23A} Arcavi, I.\ 2018, \apj, 855, 23
\bibitem[Artale et al.(2019)]{2019MNRAS.487.1675A} Artale, M. C., Mapelli, M., Giacobbo, N. et al.\ 2019, \mnras, 487, 1675
\bibitem[Bartos et al.(2017)]{2017ApJ...835..165B} Bartos, I., Kocsis, B., Haiman, Z., \& M\'arka, S.\ 2017, \apj, 835, 165
\bibitem[Banerjee et al.(2020)]{2020ApJ...901...29B} Banerjee, S., Tanaka, M., Kawaguchi, K., Kato, D., Gaigalas, G.\ 2020, \apj, 901, 29
\bibitem[Becker(2015)]{2015ascl.soft04004B} Becker, A.\ 2015, High Order Transform of PSF ANd Template Subtraction, Astrophysics Source Code Library, ascl:1504.004
\bibitem[Berger (2014)]{2014ARA&A..52...43B} Berger, E.\ 2014, \araa, 52, 43
\bibitem[Berthier et al.(2006)]{2006ASPC..351..367B} Berthier, J., Vachier, F., Thuillot, W., et al.\ 2006, in ASP Conf. Ser., 351, Astronomical Data Analysis Software and Systems XV, ed. C. Gabriel et al. (San Francisco, CA: ASP), 367
\bibitem[Bertin \& Arnouts(1996)]{1996A&AS..117..393B} Bertin, E. \& Arnouts, S.\ 1996, \aaps, 117, 393
\bibitem[Bertin et al.(2002)]{2002ASPC..281..228B} Bertin, E., Mellier, Y., Radovich, M., et al.\ 2002, ASPC, 281, 228
\bibitem[Bertin (2006)]{2006ASPC..351..112B} Bertin, E.\ 2006, Astronomical Data Analysis Software and Systems XV, 351, 112
\bibitem[Choi \& Im(2017)]{2017JKAS...50...71C} Choi, C., \& Im, M. \ 2017, JKAS, 50, 71
\bibitem[Chambers et al.(2016)]{2016arXiv161205560C} Chambers, K. C., Magnier, E. A., Metcalfe, N. et al.\ 2016, arXiv:1612.05560
\bibitem[Chornock et al.(2017)]{2017ApJ...848L..19C} Chornock, R., Berger, E., Kasen, D., et al.\ 2017, \apjl, 848, 19
\bibitem[Connaughton et al.(2016)]{2016ApJ...826L...6C} Connaughton, V., Burns, E., Goldstein, A., et al.\ 2016, \apjl, 826, 6
\bibitem[Coulter et al.(2017)]{2017Sci...358.1556C} Coulter, D. A., Foley, R. J., Kilpatrick, C. D., et al.\ 2017, Science, 358, 1556
\bibitem[Cowperthwaite et al.(2016)]{2016ApJ...826L..29} Cowperthwaite, P. S., Berger, E., Soares-Santos, M., et al.\ 2016, \apjl, 826, 29
\bibitem[Cowperthwaite et al.(2017)]{2017ApJ...848L..17C} Cowperthwaite, P. S., Berger, E., Villar, V. A., et al.\ 2017, \apjl, 848, 17
\bibitem[D\'alya et al.(2018)]{2018MNRAS.479.2374D} D\'alya, G., Galg\'oczi, G., Dobos, L., et al.\ 2018, \mnras, 479, 2374
\bibitem[de Mink \& King(2017)]{2017ApJ...839L...7D} de Mink, S. E., King, A.\ 2017, \apjl, 839, 7
\bibitem[Doctor et al.(2019)]{2019ApJ...873L..24D} Doctor, Z., Kessler, R., Herner, K., et al.\ 2019, \apjl, 873, 24
\bibitem[Drout et al.(2017)]{2017Sci...358.1570D} Drout, M. R., Piro, A. L., Shappee, B. J., et al.\ 2017, Science, 358, 1570
\bibitem[Evans et al.(2017)]{2017Sci...358.1565E} Evans, P. A., Cenko, S. B., Kennea, J. A., et al.\ 2017, Science, 358, 1565
\bibitem[Gaia Collaboration et al.(2018)]{2018A&A...616A...1G} Gaia Collaboration, Brown, A. G. A., Vallenari, A., et al.\ 2018, \aap, 616, 1
\bibitem[Goldstein et al.(2017)]{2017ApJ...848L..14G} Goldstein, A., Veres, P., Burns, E., et al.\ 2017, \apjl, 848, 14
\bibitem[Gompertz et al.(2020)]{2020MNRAS.497..726G} Gompertz, B. P., Cutter, R., Steeghs, D. et al.\ 2020, \mnras, 497, 726
\bibitem[Grado et al.(2020)]{2020MNRAS.492.1731G} Grado, A., Cappellaro, E., Covino, S. et al.\ 2020, \mnras, 492, 1731
\bibitem[Graham et al.(2020)]{2020PhRvL.124y1102G} Graham, M. J., Ford, K. E. S., McKernan, B., Ross, N. P., Stern, D. et al.\ 2020, \prl, 124, 1102
\bibitem[Hwang et al.(2021)]{2020arXiv201114049H} Hwang, S., Im, M., Taak, Y. C. et al.\ 2020, arXiv:2011.14049
\bibitem[Haggard et al.(2017)]{2017ApJ...848L..25H} Haggard, D., Nynka, M., Ruan, J. J., et al.\ 2017, \apjl, 848, 25
\bibitem[Hallinan et al.(2017)]{2017Sci...358.1579H} Hallinan, G., Corsi, A., Mooley, K. P., et al.\ 2017, Science, 358, 1579
\bibitem[Herner et al.(2020)]{2020A&C....3300425H} Herner, K., Annis, J., Brout, D., Soares-Santos, M. et al.\ 2020, A\&C, 33, 100425
\bibitem[Hotokezaka et al.(2013)]{2013ApJ...778L..16H} Hotokezaka, K., Kyutoku, K., Tanaka, M., et al.\ 2013, \apjl, 778, 16
\bibitem[Im et al.(2010)]{2010JKAS...43...75I} Im, M., Ko, J., Cho, Y., et al.\ 2010, JKAS, 43, 75
\bibitem[Im et al.(2015)]{2015JKAS...48..207I} Im, M., Choi, C., \& Kim, K.\ 2015, JKAS, 48, 207
\bibitem[Im et al.(2017)]{2017ApJ...849L..16I} Im, M., Yoon, Y., Lee, S. J., et al.\ 2017, \apjl, 849, 16
\bibitem[Im et al.(2019)]{2019JKAS...52...11I} Im, M., Choi, C., Hwang, S., et al.\ 2019, JKAS, 52, 11
\bibitem[Im et al.(2020)]{2020grbg.conf...25I} Im, M., Kim, J., Paek, G. S. -H. \& GECKO team\ 2020, Yamada Conference LXXI: Gamma-ray Bursts in the Gravitational Wave Era 2019, ed. T. Sakamoto, M. Serino \& S. Sugta. (Yokohama, Japan: Aoyama Gakuin Univ.), 25
\bibitem[Ivezi\`c et al.(2019)]{2019ApJ...873..111I} Ivezi\`c, \u{Z}., Kahn, S, M., Tyson, J. A., et al.\ 2019, \apj, 873, 111
\bibitem[Kasen et al.(2017)]{2017Natur.551...80K} Kasen, D., Metzger, B., Barnes, J., Quataert, E., Ramirez-Ruiz, E.\ 2017, \nat, 551, 80
\bibitem[Kasliwal et al.(2017)]{2017Sci...358.1559K} Kasliwal, M. M., Nakar, E., Singer, L. P., et al.\ 2017, Science, 358, 1559
\bibitem[Kim \& Im(2013)]{2013ApJ...766..109K} Kim, D., \& Im, M.\ 2013, \apj, 766, 109
\bibitem[Kim et al.(2016)]{2016JKAS...49...37K} Kim, S.-L., Lee, C.-U., Park, B.-G., et al.\ 2016, JKAS, 49, 37
\bibitem[Kim et al.(2018)]{2018JKAS...51...89K} Kim, J., Karouzos, M., Im, M., et al.\ 2018, JKAS, 51, 89
\bibitem[Lang et al.(2010)]{2010AJ....139.1782L} Lang, D., Hogg, D. W., Mierle, K., Blanton, M., \& Roweis, S.\ 2010, \aj, 139, 1782 
\bibitem[LIGO Scientific Collaboration \& VIRGO Collaboration(2019a)]{2019GCN.24069....1L} Ligo Scientific Collaboration \& VIRGO Collaboration \ 2019a, GCN, 24069, 1
\bibitem[LIGO Scientific Collaboration \& VIRGO Collaboration(2019b)]{2019GCN.24098....1L} Ligo Scientific Collaboration \& VIRGO Collaboration \ 2019b, GCN, 24098, 1
\bibitem[LIGO Scientific Collaboration \& VIRGO Collaboration(2019c)]{2019GCN.24377....1L} Ligo Scientific Collaboration \& VIRGO Collaboration \ 2019c, GCN, 24377, 1
\bibitem[Lipunov et al.(2017a)]{2017MNRAS.465.3656L} Lipunov, V. M., Kornilov, V., Gorbovskoy, E., et al.\ 2017, \mnras, 465, 3656
\bibitem[Lipunov et al.(2017b)]{2017ApJ...850L...1L} Lipunov, V. M., Gorbovskoy, E., Kornilov, V. G., et al.\ 2017, \apjl, 850, 1
\bibitem[Loeb et al.(2016)]{2016ApJ...819L..21L} Loeb, A.\ 2016, \apjl, 819, 21
\bibitem[Mapelli et al.(2018)]{2018MNRAS.481.5324M} Mapelli, M., Giacobbo, N., Toffano, M. et al.\ 2018, \mnras, 481, 5324
\bibitem[Margutti et al.(2017)]{2017ApJ...848L..20M} Margutti, R., Berger, E., Fong, W., et al.\ 2017, \apjl, 848, 20
\bibitem[McKernan et al.(2018)]{2018ApJ...866...66M} McKernan, B., Ford, K. E. S., Bellovary, J., et al.\ 2018, \apj, 866, 66
\bibitem[Morokuma et al.(2016)]{2016PASJ...68L...9M} Morokuma, T., Tanaka, M., Asakura, Y., et al.\ 2016, \pasj, 68, 9
\bibitem[Nicholl et al.(2017)]{2017ApJ...848L..18N} Nicholl, M., Berger, E., Kasen, D., et al.\ 2017, \apjl, 848, 18
\bibitem[Perna et al.(2016)]{2016ApJ...821L..18P} Perna, R., Lazzati, D., \& Giacomazzo, B.\ 2016, \apjl, 812, 18
\bibitem[Perna et al.(2018)]{2018MNRAS.477.4228P} Perna, R., Chruslinska, M., Corsi, A., \& Belczynski, K.\ 2018, \apj, 477, 4228
\bibitem[Pian et al.(2017)]{2017Natur.551...67P} Pian, E., D'Avanzo, P., Benetti, S., et al.\ 2017, \nat, 551, 67
\bibitem[Savchenko et al.(2017)]{2017ApJ...848L..15S} Savchenko, V., Ferrigno, C., Kuulkers, E., et al.\ 2017, \apjl, 848, 15
\bibitem[Smartt et al.(2016a)]{2016ApJ...827L..40S} Smartt, S. J., Chambers, K. C., Smith, K. W., et al.\ 2016, \apjl, 827, 40
\bibitem[Smartt et al.(2016b)]{2016MNRAS.462.4094S} Smartt, S. J., Chambers, K. C., Smith, K. W., et al.\ 2016, \mnras, 462, 4094
\bibitem[Smartt et al.(2017)]{2017Natur.551...75S} Smartt, S. J., Chen, T. -W., Jerkstrand, A., et al.\ 2017, \nat, 551, 75
\bibitem[Smith et al.(2019)]{2019MNRAS.485.5180S} Smith, G. P., Bianconi, M., Jauzac, M., et al.\ 2019, \mnras, 485, 5180
\bibitem[Soares-Santos et al.(2016)]{2016ApJ...823L..33S} Soares-Santos, M., Kessler, R., Berger, E., et al.\ 2016, \apjl, 823, 33
\bibitem[Soares-Santos et al.(2017)]{2017ApJ...848L..16S} Soares-Santos, M., Holz, D. E., Annis, J., et al.\ 2017, \apjl, 848, 16
\bibitem[Stalder et al.(2017)]{2017ApJ...850..149S} Stalder, B., Tonry, J., Smartt, S. J., et al.\ 2017, \apj, 850, 149
\bibitem[Stone et al.(2017)]{2017MNRAS.464..946S} Stone, N. C., Metzger, B. D., \& Haiman, Z.\ 2017, \mnras, 464, 946
\bibitem[Taak \& Im(2020)]{2020ApJ...897..163T} Taak, Y. C., \& Im, M.\ 2020, \apj, 897, 163
\bibitem[Tanaka et al.(2013)]{2013ApJ...775..113T} Tanaka, M., \& Hotokezaka, K.\ 2013, \apj, 775, 113
\bibitem[Tanaka et al.(2014)]{2014ApJ...780...31T} Tanaka, M., Hotokezaka, K., Kyutoku, K., et al.\ 2014, \apj, 780, 31
\bibitem[Tanaka et al.(2017)]{2017PASJ...69..102T} Tanaka, M., Utsumi, Y., Mazzali, P. A., et al.\ 2017, \pasj, 69, 102
\bibitem[Tanvir et al.(2017)]{2017ApJ...848L..27T} Tanvir, N. R., Levan, A. J., Gonz\'alez-Fern\'andez, C., et al.\ 2017, \apjl, 848, 16
\bibitem[Troja et al.(2017)]{2017Natur.551...71T} Troja, E., Piro, L., van Eerten, H., et al.\ 2017, \nat, 551, 71
\bibitem[Turpin et al.(2020)]{2020RAA....20...13T} Turpin, D., Wu, C., Han, X.-H. et al.\ 2017, RAA, 20, 13
\bibitem[Utsumi et al.(2017)]{2017PASJ...69..101U} Utsumi, Y., Tanaka, M., Tominaga, N., et al.\ 2017, \pasj, 69, 101
\bibitem[van Dokkum(2001)]{2001PASP..113.1420V} van Dokkum, P. G.\ 2001, \pasp, 113, 1420
\bibitem[Valenti et al.(2017)]{2017ApJ...848L..24V} Valenti, S., Sand, D. J., Yang, S., et al.\ 2017, \apjl, 848, 24
\bibitem[Onken et al.(2019)]{2019PASA...36...33O} Onken, C. A., Wolf, C., Bessell, M. S., et al.\ 2019, \pasa, 36, 33
\bibitem[Yang et al.(2019)]{2019ApJ...875...59Y} Yang, S., Sand, D. J., Valenti, S., et al.\ 2019, \apj, 875, 59
\bibitem[Yoshida et al.(2017)]{2017PASJ...69....9Y} Yoshida, M., Utsumi, Y., Tominaga, N., et al.\ 2017, \pasj, 69, 9

\end{thebibliography}
\end{document}